\documentclass[11pt,twoside]{article}

\usepackage{asp2014}

\aspSuppressVolSlug
\resetcounters

\begin{document}

\title{An Evolving Solar Data Environment}

\author{Neal Hurlburt, Sam Freeland, and Ryan Timmons}
\affil{Lockheed Martin Advanced Technology Center,  Palo Alto, CA, USA
\email{hurlburt@lmsal.com}}

\paperauthor{Hurlburt,~N.}{hurlburt@lmsal.com}{}{LMATC}{A021S}{Palo Alto}{CA}{94304 }{USA}
\paperauthor{Freeland,~S.}{freeland@lmsal.com}{}{LMATC}{A021S}{Palo Alto}{CA}{94304 }{USA}
\paperauthor{Timmons,~R.}{rtimmons@lmsal.com}{}{LMATC}{A014S}{Palo Alto}{CA}{94304 }{USA}

\begin{abstract}
The rapid growth of solar data is driving changes in the typical workflow and algorithmic approach to solar data analysis. We present recently deployed tools to aid this evolution and layout the path for future development. The majority of space-based datasets including those from the multi-petabyte Solar Dynamics Observatory and the Hinode and Interface Region Imaging Spectrograph (IRIS) missions are made available to the community through a common API with support in IDL (via SolarSoft), Python/SunPy and other emerging languages. Stellar astronomers may find the IRIS data particularly useful for research into stellar chromospheres and for interpreting UV spectra.
\end{abstract}

\section{History}
The international solar analysis environment has evolved in a bursty pattern as new missions come online.
Solar physics established a common data analysis framework based on the Interactive Data Language back in the 1990s with the support of the Yohkoh mission \citep{1991SoPh..136...37T}. Standard  FITS image metadata and a structure for integrating multiple missions and distributed, multi-lingual software trees were deployed with the support of SOHO mission \citep{1995SoPh..162....1D} creating what is now known as SolarSoft. The standardization enabled the later development of the Virtual Solar Observatory \citep{2009EM&P..104..315H}. SOHO also pushed the development of high-performance computing environments to support the computationally-intense efforts in helioseismology. 
\articlefigure{fig1.eps}{fig1}{A history of solar data analysis systems.}

World Coordinate System standards (WCS) for solar observations were developed and integrated with support of the STEREO mission \citep{2006A&A...449..791T}, while a higher-level representation of instrument coverage (including FOV, cadence, configuration) representing a coherent set of FITS files was developed under the Hinode mission. 
The Solar Dynamics Observatory \citep{2012SoPh..275....3P} pushed several efforts forward: deploying a multi-mission architecture for both data discovery, distribution and analysis. Most recently, the Interface Region Imaging Spectrograph (IRIS) has further integrated these to provide rich data descriptions and facile search interfaces. 

\section{Current Status: Infrastructure}
The current solar data environment spans the range from data discovery and exploration through delivery and analysis. It depends upon 
foundational archives and database; RESTful web services using these resources with well-defined APIs; and web tools built upon these APIs.
Here we highlight those that take advantage of the resources at LMSAL.

\subsection{Heliophysics Events Knowledgebase (HEK)}
The HEK \citep{2012SoPh..275...67H} was developed to guide solar physicists to the most relevant data for their research within the multi-petabyte SDO dataset. It automatically extracts features and events from the 1TB/day SDO data stream as it arrives and stores results in Heliophysics Event Registry (HCR). Observations from instruments with limited fields of view, such as IRIS, are  captured in the Heliophysics Coverage Registry (HCR). The HER and HCR are automatically cross-referenced to maximize the scientific discovery space. Scientists also highlight noteworthy events from their missions with our annotation tools

SolarSoft \citep{1998SoPh..182..497F} has become the foundation of the bulk of solar physics data analysis (outside of compute-intensive helioseismology studies). It is cited by nearly 1,000 referred papers since 1990 and has an h-index of 60. These papers span the domain from radio to gamma ray, solar magnetic fields and loops to eruptions and coronal mass ejections. The mostly IDL-based software tree is installed in over 6000 sites and supports 50  active and archival solar missions.

The Interface Region Imaging Spectrograph, IRIS, \citep{2014SoPh..289.2733D} is the current source of innovation in our data systems. It provides images and spectra  in the near and far ultraviolet that can be directly applied to more general astrophysics problems. The IRIS Data Search tool leverages the infrastructure provided by  SolarSoft and the HEK. IRIS observations times and fields of view are automatically cross-references with events recorded in the HEK and data from other observatories.
\section{Current Status: Webservices and Applications}
The Latest Events web pages (\url{https://www.lmsal.com/solarsoft}) fuse data from many of the active missions supported by SolarSoft.
SunToday (\url{http://sdowww.lmsal.com/suntoday_v2/index.html}) provides data discovery pages using HEK services and SDO browse products.  The new HEKsearch data search tool fully exploiting HEK services and SDO browse products while the VSO Data search fuses data from most solar missions, both active and archival. Helioviewer \citep{2009arXiv0906.1582M} is a data exploration tool using JPEG2000 compressed files that fuses data from several solar missions, including those provided by SDO. It incorporates HER services and enables citizen-science activities. 

The majority of solar data analysis and much of  science data processing are conducted using SolarSoft/IDL on local workstations and mission servers. The SolarSoft system supports both functions as part of a set of distributed software distribution tree and with a wide range of web services.
\articlefigure{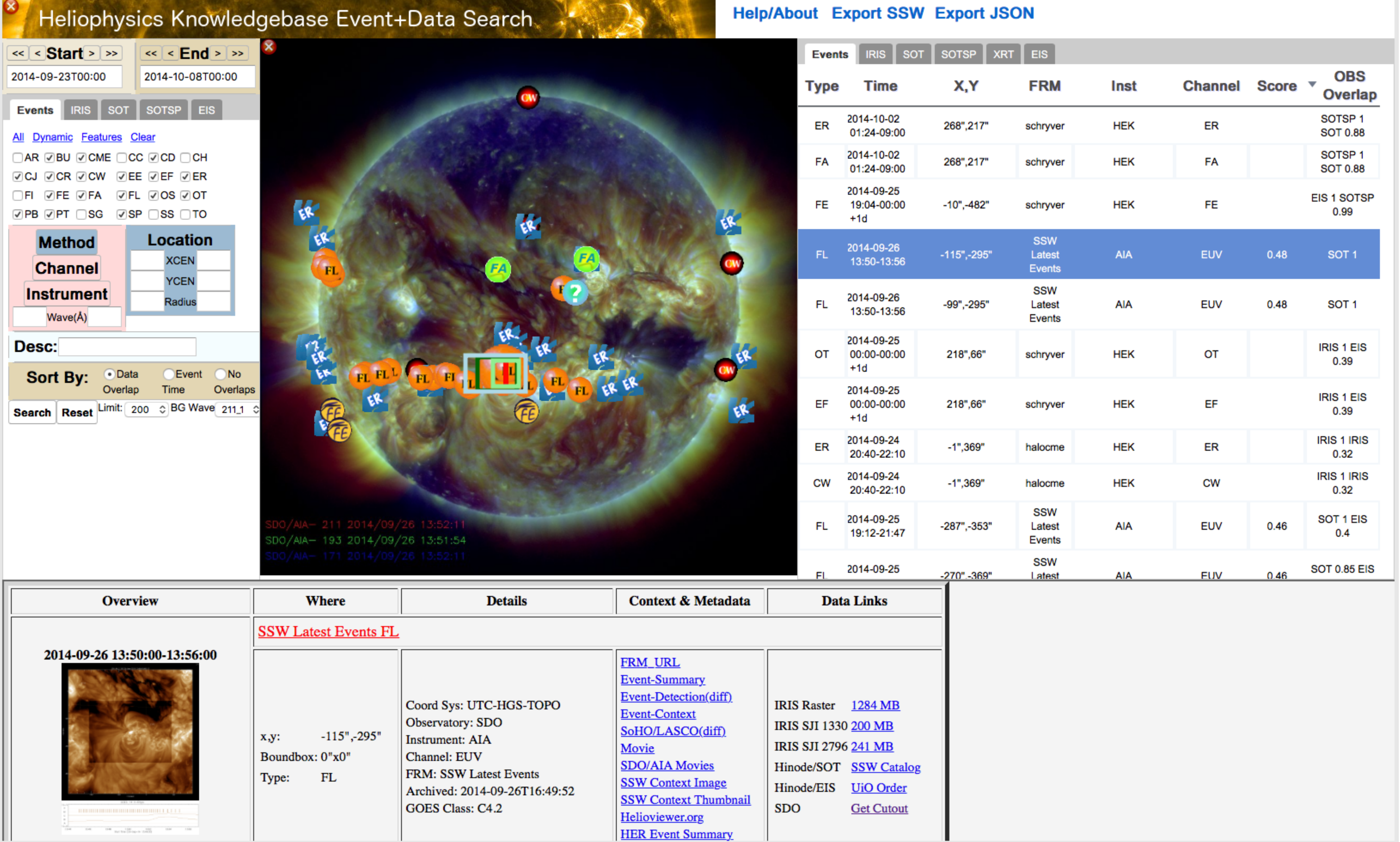}{fig2}{The HEKsearch tool provides an interface for faceted data search and discovery (\url{http://www.lmsal.com/heksearch}). }

\section{Emerging Trends}
As solar data volumes rise beyond multi-petabytes, we are closely monitoring emerging tools and technologies for adoption or incorporation into our data services and SolarSoft analysis software. We see these as emerging needs: improved data discovery; methods to move analysis codes to the data; data analytics; improving processing workflows; and improved provenance. Some tools we find promising are SunPy, Julia, and Jupyter. SunPy is the the Solar Physics package adapting AstroPy, SciPy and related packages for solar data analysis. Our future solar data processing and analysis will increasingly depend on these packages.  

\acknowledgements This work has been supported by NASA grant NNX16AB13G and contract NNG09FA40C.

\bibliography{P8-16} 

\end{document}